\documentclass{iopart}
\usepackage{graphicx}
\usepackage{amssymb}

\begin{document}

\title[Lunar Soil Strangelet Search]{Search for Stable Strange Quark Matter in Lunar Soil using the Mass Spectrometry Technique}
\author{Ke Han (for the LSSS Collaboration\footnote{The Lunar Soil
Strangelet Search Collaboration: Jeffrey Ashenfelter, Alexei
Chikanian, William Emmet, Evan Finch, Ke Han, Andreas Heinz, Richard Majka, Peter
Parker, Jack Sandweiss, Department of Physics, Yale University, New
Haven, CT 06520, USA; Benjamin Monreal, Laboratory for
Nuclear Science, Massachusetts Institute of Technology, Cambridge,
MA 02139, USA; Jes Madsen, Department of Physics and Astronomy,
University of Aarhus, DK-8000 {\AA}rhus C, Denmark})}
\address{Physics Department, Yale University, New Haven, Connecticut 06520, USA}
\ead{ke.han@yale.edu}

\submitto{\jpg}

\begin{abstract}
Strange quark matter is a postulated state which may be the true ground state of cold hadronic matter. Physicists have been searching for strange quark matter in the last several decades but found no definite evidence of its existence. In our experiment, we used the Yale tandem accelerator as a mass spectrometer to identify possible stable strangelets (small chunks of strange quark matter) in lunar soil. The search covers the mass range from A=42 to A=70 amu for nuclear charges  6, 8, and 9. No strangelets are found at sensitivity levels down to $\sim10^{-17}$.  The implied limit on strangelet flux in cosmic rays is the most sensitive to date for the covered mass range.
\end{abstract}

\section{Introduction}
Strange Quark Matter (SQM) is a hypothetical state made up with up, down and strange
quarks in a single hadronic bag~\cite{Bodmer,Witten}. Above certain baryon number threshold $A_{min}$, SQM may be energetically metastable or even absolutely stable~\cite{Farhi_Jaffe}. A simple argument can be made as follows. With baryon number $A>A_{min}$, the new flavour degree of freedom reduces the Fermi energy level $\mu_F$ of SQM compared to $ud$-quark matter. As $A$ gets larger, the benefit of an extra degree of freedom becomes so dominant that $\mu_F$ of SQM becomes even lower than that of nuclear matter. The stability of SQM has been addressed in various phenomenological models including the MIT Bag Model~\cite{Farhi_Jaffe,Chodos:1974je}. For a significant part of the "reasonable" parameter space in these models, SQM is in fact absolutely stable for $A>A_{min}$.

The signature low charge-to-mass ratio ($Z/A$) of SQM can be explained by extending the simple argument above. In a bulk system where $\mu_F$ is much greater than strange quark mass $m_s$ ($\mu_F\gg m_s$), the numbers of three quark flavours are almost identical ($N_u=N_d\simeq N_s$). Therefore, bulk SQM is almost electric charge neutral. The system can be arbitrarily huge (as large as a star) without fission because of the charge neutrality and the consequential minimal Coulomb repelling effects. SQM with small baryon number and $\mu_F\gtrsim m_s$, usually called a strangelet, is positively charged because of strange quark number deficit ($N_u=N_d> N_s$). Still, the charge-to-mass ratio of strangelets is much smaller than normal nuclear matter. According to the MIT bag model calculation~\cite{Farhi_Jaffe,Chodos:1974je}, $Z=0.1A$ for $A<700$ amu. Bag model calculation with colour-flavour-locking in consideration~\cite{ Alford:1998mk,Madsen:2001fu} shows that $Z=0.3A^{2/3}$. Both formulae assume $m_s=150$ MeV.

Strangelets are usually categorized according to their nuclear charges and baryon numbers, the same way as normal isotopes. For example, a strangelet with charge $Z=+6$, mass $A=60$ amu is called strange Carbon and simply denoted as $^{60}\mathrm{C}$.

If completely stable at zero pressure, then SQM is the true ground state of cold hadronic matter~\cite{Witten}. The implication of this hypothesis is tremendous for many aspects of physics~\cite{Shaw:1988pc}. One of the most striking consequences is that all compact stars are in fact SQM's (usually called "strange stars"), not neutron star~\cite{Witten,StrangeStars}. Strangelets, which are ejected in the binary strange star collisions, experience various acceleration processes in the universe and ultimately reach the solar system together with other cosmic ray particles~\cite{MedinaTanco:1996je,Madsen:2004vw}.

The moon is a much better depository for cosmic strangelets than the Earth because of its lack of geological activities, atmosphere, or geomagnetic field. We obtained from NASA a 15 g sample of lunar soil from sample No. 10084.  This fine particulate sample was collected from the top 7.5 cm of the lunar surface~\cite{Apollo11PSR} and has a cosmic rays exposure age of $520\pm 120$ Myr~\cite{SoilAge}. For strangelets with mass less than 100 amu, the abundance in the lunar soil sample is about 1 part per $10^{15}\sim10^{16}$, calculated based on theoretical cosmic strangelet flux estimation.

Candidate events consistent with strangelet characteristics have been reported by several experiments. Three different balloon-borne experiments published four candidate events with long stopping ranges and small $Z/A$~\cite{Price_ET,Saito_ET,Ichimura_ET}. Two other strangelet candidate events were identified by the Alpha Magnetic Spectrometer (AMS) collaboration during the AMS-01 prototype flight in 1998~\cite{Aguilar:2002ad}. One of the events was reconstructed as $^{54}\mathrm{O}$ with $A=54^{+8}_{-6}$ and the other event as $^{16}\mathrm{He}$~\cite{AMS_He}.

Our experiment, which aims to confirm or rule out the AMS-01 $^{54}\mathrm{O}$ event, searches for low mass ($A=42\sim70$ amu, $2\sigma$ range of the $^{54}\mathrm{O}$ event) strangelet relics in lunar soil using the tandem Van de Graaff accelerator at Yale A. W. Wright Nuclear Structure Laboratory (WNSL)~\cite{WNSL_Tandem88} as a mass spectrometer. The search reaches single event sensitivity levels around $3\times10^{-17}$ and the implied sensitivity to SQM as a component of cosmic rays falls below both the theoretical flux prediction and the flux implied by the AMS-01 $^{54}\mathrm{O}$ candidate event.

\section{Experimental Setup}
\begin{figure}[t]
\begin{flushright}
\includegraphics[height=0.8\textwidth, angle=90]{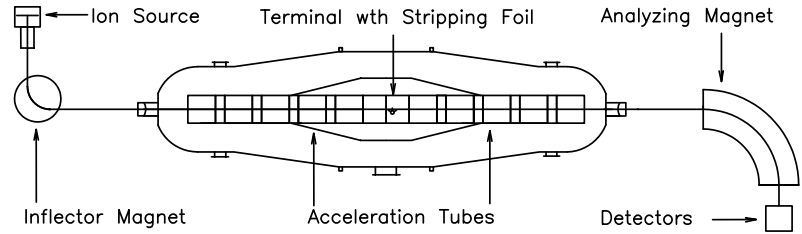}
\caption{\label{fig:apparatus} Schematic figure of the Yale WNSL Tandem
accelerator (not to scale). Beam direction is from left to right.}
\end{flushright}
\end{figure}

The Yale Tandem is shown schematically in Figure~\ref{fig:apparatus} with several key components highlighted. Lunar soil sample (used in 0.1g divisions) is pressed into the ion source cone, where negative ions are formed by Cesium ion sputtering~\cite{IonSource_Middleton83}. Negative ions are pre-accelerated from ground to 20KV for the mass selection in the inflector magnet. The magnet is set to transmit ions with a given mass $A_0$ and charge $Q=-1$. Following this, the ions enter the main acceleration tank and are accelerated to the positive terminal at 17 MV in the middle, where a 10 $\mu g / cm^2$ Carbon foil strips electrons from the ions.  For a $^{54}\mathrm{O}$ strangelet with an incident energy of 17 MeV, the most abundant charge state emerging from the foil is $Q=+5$~\cite{Shima:1986}. The stripped ions are then accelerated away from the terminal to ground, and then go through an analyzing magnet in which ions with mass $A_0$, charge $Q=+5$, and total energy 102 MeV survive before entering our detector system. The mass acceptance $\delta_m / m$ of the inflector and analyzing magnets (with all slits wide open) are both about 0.6\% so that a mass range of $\delta_m = 1/4$ amu can be covered in each run.

The performance of the accelerator is monitored with extra caution. Electrostatic accelerators usually rely on ion beam feedback to regulate the terminal voltages. However, in our experiment, when set for a mass $A_0$ which is not an integer, there is no normal nuclear ion beam transmitted through the machine, and thus no feedback. Therefore, terminal voltage in this experiment is held constant by a feedback system utilizing a set of Generating Voltmeters inside the accelerator tank wall. The stability is ensured by periodic short checks roughly every 4 hours.

\begin{figure}[t]
\begin{flushright}
\includegraphics[width=0.8\textwidth]{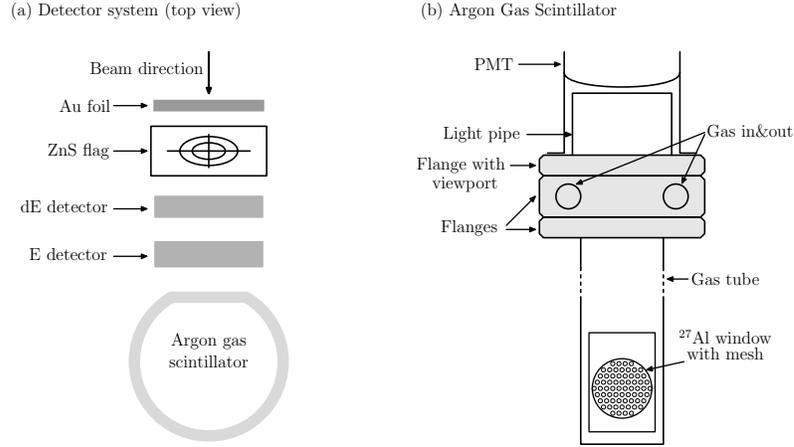}
\caption{\label{fig:detectors} (a). Schematic setup of the detector system. All the components, except the Argon gas scintillator, can be removed from and inserted into the beam line remotely. (b). Schematic plot of the Argon gas scintillator.}
\end{flushright}
\end{figure}

Our experiment exploits strangelets' characteristic small charge-to-mass ratio and the resulting long dE/dx stopping range to identify any strangelet event. The detector system, including a 10 $\mu m$ thick Gold foil, a ZnS fluorescent screen, two silicon detectors, and a gaseous Argon scintillation counter, is shown schematically in Figure~\ref{fig:detectors}(a). All the components, except the scintillation counter, can be withdrawn from or inserted into the beam line remotely.

For normal running during the strangelet search, the gold foil and both silicon detectors are inserted into the beamline. The first silicon detector (dE detector), which measures the energy loss of a penetrating strangelet, has a circular cross-sectional active area of 40 $mm^2$ and is about $12\, \mu m$ thick.  The second silicon detector (E detector), which measures the remaining energy, has a cross section of 100~$mm^2$ and is 100~$\mu m$ thick. Strangelets with 102 MeV incident energy penetrate the foil and leave well defined signals in the two silicon detectors~\cite{SRIMbook}. For example, a $^{54}\mathrm{O}$ ion loses about 60 MeV in the foil; deposits about 16 MeV in the dE detector and leaves the remaining 24 MeV for collection by the E detector. On the contrary, a normal nucleus of comparable mass and incident energy stops inside the Gold foil.

The Argon scintillation counter and the ZnS screen (viewed by a camera imaging the screen) are used as monitors of beam quality and stability throughout the running period and to make various transmission measurements. The scintillation counter is designed and built in-house (see Figure~\ref{fig:detectors}(b)). Argon gas flows inside a stainless steel tube with front aluminium window of 1 cm in diameter and only 6 $\mu m$ thick. To hold the one atmosphere pressure difference in the gas tube and the accelerator vacuum, the window is supported by stainless steel mesh with about 43\% open area. The inner wall of the tube is firstly painted with a reflective layer and then coated with tetraphenyl butadiene (TPB)~\cite{TPBcoating} to shift wavelength from 128nm~\cite{ArgonScint} to 440nm (blue), which is better matched to the response of the phototube. With our design, the gas scintillation counter can withstand a counting rate about 1 MHz. By comparison, the plastic scintillator which we used initially started yellowing after being exposed to several kHz of beam.

\section{Results}
\begin{figure}
\begin{flushright}
\includegraphics[width=0.8\textwidth]{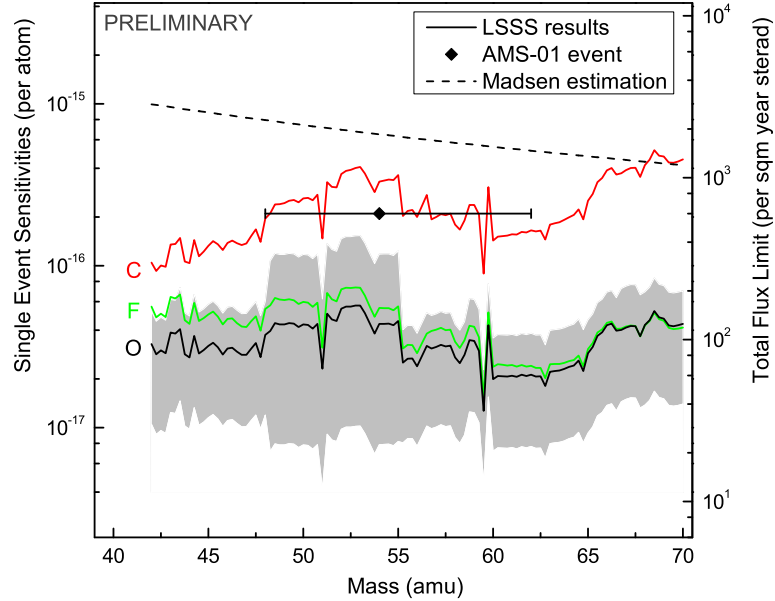}
\caption{\label{fig:lsss_results} Single event sensitivity of our
strangelet search as a function of nuclear mass for various values
of nuclear charge.  The Y axis on the right shows the implied flux
limits for cosmic ray strangelets.}
\end{flushright}
\end{figure}

We have searched over a mass range from 42 to 70 amu and found no strangelet candidate events in this experiment. Some background particles, most abundantly knocked-out Carbon ions with non-rational energies, may survive the analyzing magnet's rigidity selection and enter our detector system. But in no case were these signals within 10 MeV of the expected strangelet signal on the dE/E plot. The experiment was therefore free of background. The single event sensitivity limit for strange Oxygen with respect to normal Oxygen with a given mass setting $A_0$ can be calculated as
\begin{equation}
s=\frac{1}{I \times T \times P_{+5} \times \epsilon_T(5) } \label{eq:sensitivity}
\end{equation}
where $I$, the current of $^{16}\mathrm{O}$  out of the ion source averaged approximately $7\times10^{13}$ particles per second. $T$, the running time per mass setting $A_0$, was nominally two hours. $P_{+5}=40\%\sim50\%$~\cite{Shima:1986} is the probability of a strangelet Oxygen being stripped to a charge state of $Q=+5$ in the Carbon stripping foil. $\epsilon_T(5)$, the transmission efficiency of charge +5 beam through the tandem, varied from about 4\% to 9\% in different runs. The sensitivity with respect to all normal atoms in lunar soil (shown as a solid black line in Figure~\ref{fig:lsss_results}) is about $s/1.5$, considering the relative abundance of Oxygen atoms in lunar soil and negative ion forming efficiency by sputtering.

The dominant systematic uncertainty in our sensitivity comes from the determination of $\epsilon_T(5)$.  Direct measurement of $\epsilon_T(5)$ was unfeasible because of the large uncertainty of stripping probability $p_{+5}$ for a normal nucleus like Nickel with an incident energy of 17 MeV. We determined $\epsilon_T(5)$ by measuring $\epsilon_T(Q)$ vs. stripped charge $Q$ for a variety of mass states to determine the dependence of $\epsilon_T(Q)$ on mass and charge. From the reproducibility of and variation in these measurements, we estimated a systematic uncertainty of $\pm50\%$ to $\epsilon_T(5)$.

The search was optimized for strange Oxygen, but was also sensitive to strangelets of different $Z$ values. The search limits are different than the $Z=8$ limits because nuclei of other $Z$ values generally have different efficiencies for producing negative ions in the sputtering ion source~\cite{Middleton1989} and different probabilities for stripping to $Q=+5$ in the Carbon foil at the terminal~\cite{Shima:1986}. When these differences are accounted for, we obtain the limits for strange Carbon and Fluorine shown in Figure~\ref{fig:lsss_results}. This experiment is not sensitive to strange Nitrogen or Neon at all because these elements do not form negative ions by sputtering.

These sensitivity results can be transformed into limits on the strangelet flux based on the lunar sample's exposure age to cosmic rays ($520\pm120$ Myr) and the strangelets' deposition depth in lunar soil. For a commonly used energy spectrum~\cite{Madsen:2004vw}, approximately 40\% of the strangelets under study here would stop in the top 7.5 cm of lunar material, where the lunar soil sample No. 10084 was collected. Using these numbers, the upper flux limits determined from this search are shown in Figure~\ref{fig:lsss_results} on the right Y axis. For comparison, the AMS-01 $^{54}\mathrm{O}$ candidate and the theoretical predictions by Madsen~\cite{Madsen:2004vw} are also shown in the figure.

\section{Conclusion and Outlook}
The important result of this search is that the AMS-01 Z=8 candidate event is probably ruled out. We will follow up with detailed analysis about systematic uncertainties very soon. The flux limit for $Z=8$, $42<A<70$ strangelets in cosmic rays determined here is some 4 orders of magnitude below previous searches(see e.g.~\cite{Finch:2006pq} for a review of previous searches). We also show the best existing limit for nearby charge states like strange Carbon and Fluorine.

If the AMS-02 experiment is launched onto the ISS in the near future, it will be an almost definitive search for cosmic ray strangelets (and, by extension, a very strict test of the hypothesis of stable SQM).  In the unlikely event that the AMS-02 is further postponed or cancelled, there is still room for improvement in searches such as the one reported here.  We can extend our coverage towards higher mass range (80 to 140 amu) which is considered theoretically to be the more likely (It's worth noting that shell model effects are expected to be quite important at such low masses so that these ranges should be seen as very rough guidelines).  Also, we could improve our limits drastically by enriching the heavy isotope concentration in the lunar samples, though the chemical enriching process completely eliminates elements of different $Z$ value other than Oxygen and thus restricts our sensitivity to strange Oxygen only.

In conclusion, the existence of stable strange quark matter remains an open question. Physicists will keep searching.

\section*{Acknowledgements}
The LSSS Collaboration gratefully acknowledges very valuable work by Thomas Hurteau, John Baris, Thomas Barker, Walter Garnett and other WNSL operators. They also thank Richard Casten for generously providing us accelerator beam time at WNSL. The work is supported by US Department of Energy under contract No. DE-FG02-92ER40704.


\section*{References}
\begin{thebibliography}{10}
\providecommand{\url}[1]{\texttt{#1}}
\providecommand{\urlprefix}{URL }

\bibitem{Bodmer}
Bodmer A~R 1971 \emph{Phys. Rev. D} \textbf{4} 1601

\bibitem{Witten}
Witten E 1984 \emph{Phys. Rev. D} \textbf{30} 272

\bibitem{Farhi_Jaffe}
Farhi E and Jaffe R~L 1984 \emph{Phys. Rev. D} \textbf{30} 2379

\bibitem{Chodos:1974je}
Chodos A \emph{et~al.} 1974 \emph{Phys. Rev. D} \textbf{9} 3471

\bibitem{Alford:1998mk}
Alford M~G, Rajagopal K and Wilczek F 1999 \emph{Nucl. Phys. B} \textbf{537}
  443

\bibitem{Madsen:2001fu}
Madsen J 2001 \emph{Phys. Rev. Lett.} \textbf{87} 172003

\bibitem{Shaw:1988pc}
Shaw G~L \emph{et~al.} 1989 \emph{Nature} \textbf{337} 436

\bibitem{StrangeStars}
Alcock C, Farhi E and Olinto A 1986 \emph{Astrophys. J.} \textbf{310} 261

\bibitem{MedinaTanco:1996je}
Medina-Tanco G~A and Horvath J~E 1996 \emph{Astrophys. J.} \textbf{464} 354

\bibitem{Madsen:2004vw}
Madsen J 2005 \emph{Phys. Rev. D} \textbf{71} 014026

\bibitem{Apollo11PSR}
NASA 1969 \emph{Apollo 11 Preliminary Science Report} unpublished

\bibitem{SoilAge}
{Eberhardt} P \emph{et~al.} 1970 \emph{Geochim. Cosmochim. Acta Suppl.}
  \textbf{1} 1037

\bibitem{Price_ET}
Price P~B \emph{et~al.} 1978 \emph{Physical Review D} \textbf{18} 1382

\bibitem{Saito_ET}
Saito T \emph{et~al.} 1990 \emph{Physical Review Letters} \textbf{65} 2094

\bibitem{Ichimura_ET}
Ichimura M \emph{et~al.} 1993 \emph{Nuovo Cimento A} \textbf{106} 843

\bibitem{Aguilar:2002ad}
Aguilar M \emph{et~al.} 2002 \emph{Phys. Rept.} \textbf{366} 331

\bibitem{AMS_He}
Choutko V 2003 in \emph{28th Int. Cosmic Ray Conf} 1765 Tsukuba, Japan

\bibitem{WNSL_Tandem88}
Hyder H~R~M \emph{et~al.} 1988 \emph{Nucl. Instrum. Methods A} \textbf{268} 285

\bibitem{IonSource_Middleton83}
Middleton R 1983 \emph{Nucl. Instrum. Methods} \textbf{214} 139

\bibitem{Shima:1986}
Shima K, Mikumo T and Tawara H 1986 \emph{At. Data Nucl. Data Tables}
  \textbf{34} 357

\bibitem{SRIMbook}
Ziegler J~F \emph{et~al.} 1985 \emph{The stopping and range of ions in solids}
  Pergamon, New York

\bibitem{TPBcoating}
Burton W~M and Powell B~A 1973 \emph{Appl. Opt.} \textbf{12} 87

\bibitem{ArgonScint}
Policarpo A~J~P~L 1981 \emph{Physica Scripta} \textbf{23} 539

\bibitem{Middleton1989}
Middleton R 1989 \emph{A Negative Ion Cookbook} Unpublished

\bibitem{Finch:2006pq}
Finch E 2006 \emph{J. Phys. G} \textbf{32} S251

\end{thebibliography}
\end{document}